\documentclass[aps,pra,twocolumn,nofootinbib]{revtex4}
\usepackage{amsmath,amssymb}

\newcommand{\cH}{\mathcal{H}}
\newcommand{\cE}{\mathcal{M}}
\newcommand{\Tr}{\mathrm{Tr}}
\newcommand{\bX}{\mathbf{X}}
\newcommand{\bP}{\mathbf{P}}

\newcommand{\bZ}{\mathbf{Z}}
\newcommand{\Th}{\mathrm{th}}

\newcommand{\reals}{\mathbb{R}}

\newcommand{\pmat}[1]{\ensuremath{\begin{pmatrix}#1\end{pmatrix}}}
\newcommand{\ket}[1]{\ensuremath{\left| #1\right\rangle}}
\newcommand{\bra}[1]{\ensuremath{\left\langle #1\right|}}
\newcommand{\braket}[2]{\ensuremath{\left\langle #1|#2\right\rangle}}
\newcommand{\ketbra}[2]{\ket{#1}\!\!\bra{#2}}
\newcommand{\expect}[1]{\ensuremath{\left\langle#1\right\rangle}}

\newcommand{\proj}[1]{\ketbra{#1}{#1}}
\def\Id{1\!\mathrm{l}}
\newcommand{\Avg}{\mathop{\mathrm{Avg}}}
\newcommand{\diff}{\mathrm{d}\!}

\begin{document}
\author{Robin Blume-Kohout}
\affiliation{Theoretical Division, Los Alamos National Laboratory; Los Alamos, NM 87545}
\author{Peter S. Turner}
\affiliation{Department of Physics, Graduate School of Science, The University of Tokyo, 7-3-1 Hongo, Bunkyo-ku, Tokyo, Japan 113-0033}

\title{The curious nonexistence of Gaussian 2-designs}


\begin{abstract}
2-designs -- ensembles of quantum pure states whose 2nd moments equal those of the uniform Haar ensemble -- are optimal solutions for several tasks in quantum information science, especially state and process tomography.  We show that Gaussian states cannot form a 2-design for the continuous-variable (quantum optical) Hilbert space $L^2(\reals)$.  This is surprising because the affine symplectic group HWSp (the natural symmetry group of Gaussian states) is irreducible on the symmetric subspace of two copies.  In finite dimensional Hilbert spaces, irreducibility guarantees that HWSp-covariant ensembles (such as mutually unbiased bases in prime dimensions) are always 2-designs.  This property is violated by continuous variables, for a subtle reason:  the (well-defined) HWSp-invariant ensemble of Gaussian states does not have an average state because the averaging integral does not converge.  In fact, no Gaussian ensemble is even \emph{close} (in a precise sense) to being a 2-design.  This surprising difference between discrete and continuous quantum mechanics has important implications for optical state and process tomography.
\end{abstract}
\maketitle

A quantum $t$-design is an ensemble of pure quantum states, whose $t^\Th$ (and lower) moments mimic those of the unitarily invariant Haar ensemble, which captures the notion of `random' pure states in Hilbert space.  Designs have a variety of applications in quantum information science and the foundations of quantum theory, most notably in quantum \emph{tomography} \cite{WoottersAP89,ScottJPA06,GrasslENDM05,MedendorpPRA11}.  To date, designs have been used primarily in finite dimensional Hilbert spaces, with optical coherent states \cite{GlauberPRA63,Perelomov86,LvovskyRMP09} (a 1-design for continuous variables) as an outstanding exception.  While 1-designs are useful, for example as resolutions of the identity operator, 2-designs seem to be the most useful and interesting of designs.  They are far superior to 1-designs -- often optimal -- for a variety of tasks, including quantum state and process tomography \cite{ScottJPA06,ScottJPA08}, and maximal unclonability \cite{Fuchs03}.  Relatively few applications \cite{AmbainisCCC07,HarrowLNCS09} are known for higher-order designs.  Finite-dimensional 2-designs include mutually unbiased bases (MUBs) (see, e.g. \cite{WoottersAP89,GibbonsPRA04}) and symmetric informationally complete positive operator valued measures (SICPOVMs) \cite{RenesJMP04}.

In this paper, we attempt to construct a continuous-variable 2-design from Gaussian states (including coherent states \emph{and} their squeezed cousins).  There is ample reason to suspect that such a construction is possible.  The simplest 2-design construction in $d$-dimensional finite Hilbert spaces comprises $(d+1)$ mutually unbiased bases in prime power dimensions ($d = p^n$) \cite{WoottersAP89}.  The basis states' discrete Wigner functions form lines in discrete phase space.  These constructions, and indeed the whole idea of discrete phase space, were motivated by continuous-variable phase space and the associated Hilbert space $L^2(\reals)$.

Thus, we are surprised to report that Gaussian 2-designs do \emph{not} exist.  At first inspection (Section \ref{sec:sp}), it appears that they should because their natural transitive symmetry group -- the affine symplectic group -- acts irreducibly on the symmetric subspace of $L^2(\reals)\otimes L^2(\reals)$.   Schur's Lemma implies that a symplectically-invariant ensemble of Gaussians should therefore form a 2-design.  However, an explicit calculation of matrix elements (Section \ref{sec:no}) shows that it is impossible to construct a 2-design out of Gaussian states.  In fact, it's impossible even to get close!  The resolution (Section \ref{sec:discuss}) involves the non-convergence of an integral over the symplectic group.  Because this integral fails to converge, a symplectically-invariant mixture of Gaussian states does not exist -- which is a sort of end run around Schur's Lemma.  We point out that this is not merely due to the noncompactness of the symplectic group, for integrations over noncompact groups are often well-defined and yield invariant quantities (as demonstrated by the coherent states).

\section{Designs and representations} \label{sec:background}

A set of states $\cE$ on a Hilbert space $\cH$ is a $t$-design for $\cH$ if its $t^\Th$ moments are identical to those of the unitarily invariant (Haar) ensemble of pure states on $\cH$:
\begin{equation}
\Avg_{\psi\in\cE}\left(\proj{\psi}^{\otimes t}\right) = \int_{\psi\in\mathrm{Haar}}{\proj{\psi}^{\otimes t}\diff\psi}. \label{eq:tdesign1}
\end{equation}
The Haar average on the R.H.S. of Eq. \ref{eq:tdesign1} can be calculated easily using Schur's Lemma, which states:  if a nonzero operator on an irreducible representation (irrep) space of a group $G$ commutes with every element of that irrep, it must be proportional to $\Id$ on that space.  The R.H.S. commutes (by construction) with every $U^{\otimes t}$, so it must be a sum of irrep projectors.  Since $\proj{\psi}^{\otimes t}$ lies entirely in the symmetric subspace $\cH_{\mathrm{symm}}^{(t)}$ of $\cH^{\otimes t}$ (which is an irrep space), the Haar average is proportional to the projector onto $\cH_{\mathrm{symm}}^{(t)}$, and the $t$-design condition is
\begin{equation}
\Avg_{\psi\in\cE}\left(\proj{\psi}^{\otimes t}\right) = \frac{\Pi_{\mathrm{symm}}^{(t)}}{\Tr\left(\Pi_{\mathrm{symm}}^{(t)}\right)}. \label{eq:tdesign}
\end{equation}

Here we are concerned with Gaussian states on the Hilbert space $L^2(\reals)$, and whether it is possible to construct a 2-design from them.  The coherent states are a simple example of a Gaussian 1-design, and we will review this construction in order to introduce some simple, useful group theory tricks.

\section{Heisenberg-Weyl covariant 1-designs} \label{sec:h}

Schur's Lemma applies to any group.  This suggests an elegant way to construct designs.  Choose a group $G$ that has a unitary representation $T$ on $\cH$,
\begin{equation*}
g\to T_g, \quad g\in G.
\end{equation*}
If the natural $t$-copy tensor product representation
\begin{equation*}
g\to T_g^{\otimes t}
\end{equation*}
is irreducible on the symmetric subspace\footnote{Note that the $\{T_g^{\otimes t}\}$ representation commutes with permutations of the $t$ copies, so its action on on $\cH^{\otimes t}$ is reducible onto the irrep spaces of the symmetric group $S_t$. $\cH_{\mathrm{symm}}^{(t)}$ is one of these.} of $\cH^{\otimes t}$, then Schur's Lemma implies that the ensemble
\begin{equation*}
\cE = \{T_g\ket{\psi_0}\ \forall\ g\in G\}
\end{equation*}
is a $t$-design for any $\ket{\psi_0}\in\cH$.

$\cE$ will be a 1-design if $T_g$ itself is irreducible on $\cH$.  For example, the Heisenberg-Weyl group on one degree of freedom, HW$(1)$ (or simply HW, hereafter), has an irreducible representation on $\cH = L^2(\reals)$ as translations in phase space (displacement operators)
\begin{equation*}
\{T_{x,p,\phi}\} \equiv \{e^{-i(\phi + x\bP + p\bX)/\hbar}\ \forall\ x,p\in\reals\ \mathrm{and}\ \phi\in[0,2\pi]\},
\end{equation*}
where $\bX$ and $\bP$ are the position and momentum operators on $L^2(\reals)$.  HW is a noncompact Lie group.  Its natural Haar measure is $\mu = \diff x\diff p\diff\phi$ -- which is simply the Lebesgue measure over phase space and over the phase $\phi$.  Because we use state \emph{projectors} $\proj{\psi}$ exclusively, central phases such as $\phi$ vanish.  We can safely treat all group representations as projective, and will denote this phase-space-translation representation of the Heisenberg-Weyl group by $\{T_{x,p}\}$.

Since $\{T_{x,p}\}$ is irreducible on $L^2(\reals)$, any ``fiducial state'' $\ket{\psi_0}$ can be used to generate a Heisenberg-Weyl (HW)-covariant 1-design 
\begin{equation*}
\cE_1 = \{T_{x,p}\ket{\psi_0}\},
\end{equation*}
weighted according to the invariant measure $\mu$ of its defining group.  The coherent state POVM (positive operator valued measure\footnote{A pedagogical note on the term ``POVM'' may be useful here.  A POVM is a measure over a sample space; the sample space is the set of possible outcomes of a quantum measurement.  To each [measurable] subset of outcomes, the POVM assigns a positive operator (whereas conventional measures would assign a positive number).  Often, in quantum information science, the sample space is finite -- and so measure theory is mostly superfluous, and practitioners often forget just what ``POVM'' means.  Here, our sample space is the continuously infinite manifold of a Lie group, and [basic] measure theory is required.}) of quantum optics is generated by choosing $\ket{\psi_0}$ to be the ``vacuum'' state $\ket{0}$ of a dimensionless ($m=\omega=\hbar=1$) harmonic oscillator Hamiltonian $H = \bX^2 + \bP^2$, with wavefunction
\begin{equation*}
\psi_0(x) = \braket{x}{\psi_0} = \frac{e^{-x^2/2}}{(\pi)^{1/4}}.
\end{equation*}

The coherent states are not unique; we can generate a HW-covariant Gaussian 1-design from any Gaussian fiducial $\ket{\psi_0}$.  But none of them are 2-designs, and the proof is rather instructive.

\section{Heisenberg-Weyl is insufficient for 2-designs}

To evaluate whether an ensemble $\cE$ is a 2-design, we consider 
\begin{equation*}
\cH^{\otimes 2} = \cH\otimes\cH = L^2(\reals)\otimes L^2(\reals),
\end{equation*}
containing square-integrable functions $\psi(x_1,x_2)$ of \emph{two} real variables.  For any group $G$, the tensor product representation $T_g\otimes T_g$ is reducible onto the symmetric and antisymmetric subspaces of $\cH^{\otimes 2}$, because $T_g\otimes T_g$ commutes with the SWAP operator $\pi$ that permutes the two systems.  The SWAP operator becomes very simple if we change coordinates from $(x_1,x_2)$ to $\left( x_+ = \frac{x_1+x_2}{\sqrt2}, x_- = \frac{x_1-x_2}{\sqrt2} \right)$, which defines a refactorization of the Hilbert space as
\begin{equation*}
L^2(x_1)\otimes L^2(x_2) \to L^2(x_+)\otimes L^2(x_-).
\end{equation*}
Now SWAP has no effect on the $x_+$ subsystem (since $x_2+x_1 = x_1+x_2$), but acts on the $x_-$ subsystem as the parity operator $P:x\to -x$, because $x_2-x_1 = -(x_1-x_2)$.  The projectors onto the symmetric and antisymmetric subspaces are thus:
\begin{eqnarray*}
\Pi_{\mathrm{symm}}^{(2)} &=& \Id_{+}\otimes\frac{(\Id + P)_-}{2} \\
\Pi_{\mathrm{antisymm}}^{(2)} &=& \Id_{+}\otimes\frac{(\Id - P)_-}{2}
\end{eqnarray*}
But the HW representation $\{T_{x,p}^{\otimes 2}\}$ is \emph{not} irreducible on these subspaces.  Its elements act on $L^2(x_+)\otimes L^2(x_-)$ as
\begin{eqnarray*}
T_{x,p}\otimes T_{x,p} 
 &=& e^{-i(x\bP_1 + x\bP_2 + p\bX_1 + p\bX_2)/\hbar} \\
 &=& e^{-i\sqrt2(x\bP_+ + p\bX_+)/\hbar}\otimes\Id_-,
\end{eqnarray*}
so they act faithfully on $L^2(x_+)$, but trivially on $L^2(x_-)$.  \emph{Any} 1-dimensional subspace of $L^2(x_-)$ -- e.g., $\mathrm{span}\{\ket{\psi}_-\}$ -- is an irrep space of this representation,
\begin{equation*}
L^2(x_+) \otimes \mathrm{span}\{\ket{\psi}_-\}.
\end{equation*}
Note that this structure is slightly different from the usual ``direct sum of irrep spaces'' structure.  As a direct sum of uncountably many copies of the irrep $\{T_{\sqrt{2}x,\sqrt{2}p}\}$, it is much more naturally described as a tensor product.

This simple representation structure makes it easy to evaluate whether a Heisenberg-Weyl-covariant ensemble is a 2-design.  We tensor two copies of the fiducial state $\ket{\psi_0}$ and rewrite it in the refactored Hilbert space using its Schmidt decomposition:
\begin{equation*}
\ket{\psi_0}\ket{\psi_0} = \sum_k{ c_k \ket{k}_+\otimes \ket{k}_- },
\end{equation*}
where $\ket{k}_\pm$ are elements of orthonormal bases for $L^2(x_\pm)$.
This state is generally entangled, with more than one Schmidt coefficient $c_k$.  Now, the average of $\proj{\psi}^{\otimes 2}$ is equal to the averaged action of the Heisenberg-Weyl group on $\proj{\psi_0}^{\otimes 2}$,
\begin{equation*}
\Avg_\cE\left( \proj{\psi}^{\otimes 2} \right) = \Avg_{\mathrm{HW}}\left(T_{x,p}^{\otimes 2}\proj{\psi_0}^{\otimes 2}\left(T^\dagger_{x,p}\right)^{\otimes 2}\right)
\end{equation*}
Since HW acts irreducibly on $L^2(x_+)$, it will completely depolarize the $L^2(x_+)$ subsystem, and destroy all correlation with the $L^2(x_-)$ subsystem:
\begin{equation*}
\Avg_\cE\left( \proj{\psi}^{\otimes 2} \right) \propto \Id_+ \otimes \sum_k{ |c_k|^2 \proj{k}_- }.
\end{equation*}
This is proportional to $\Pi_{\mathrm{symm}}^{(2)}$ if and only if $\ket{\psi_0}\ket{\psi_0}$ is maximally entangled between $L^2(x_+)$ and the positive-parity subspace of $L^2(x_-)$.  Such states exist for many (perhaps all) \emph{finite}-dimensional Hilbert spaces, which admit a discrete analogue of the Heisenberg-Weyl group, but finding them is an open challenge known as the SICPOVM problem \cite{RenesJMP04}.  

But when $\ket{\psi_0}$ is a Gaussian state, $\ket{\psi_0}\ket{\psi_0}$ is actually a product state of $L^2(x_+)$ and $L^2(x_-)$.  For a completely general Gaussian wavefunction
\begin{equation*}
\psi_0(x) \propto e^{-(\alpha+i\beta)x^2 + (\gamma+i\delta)x},
\end{equation*}
the wavefunction for $\ket{\psi_0}\ket{\psi_0}$ is (in the $x_{\pm}$ coordinates)
\begin{eqnarray*}
\psi_0^{\otimes2}(x_+,x_-) 
&=& \psi_0\left(\frac{x_+ + x_-}{\sqrt2}\right)\psi_0\left(\frac{x_+ - x_-}{\sqrt2}\right) \\ \nonumber
&\propto & e^{-(\alpha+i\beta)x_+^2 + \sqrt2(\gamma+i\delta)x_+} \cdot e^{-(\alpha+i\beta)x_-^2}.
\end{eqnarray*}
So $\psi_0^{\otimes2}(x_+,x_-) =: \psi_+(x_+)\psi_-(x_-)$ is a product state, and 
\begin{equation*}
\Avg_\cE(\proj{\psi}^{\otimes 2}) \propto \Id_+ \otimes \proj{\psi_-} \neq \Pi_{\mathrm{symm}}^{(2)}.
\end{equation*}
In fact, the two-copy average state is about as far as possible from the projector on the symmetric subspace, since it is rank-1 on $L^2(x_-)$.  The Heisenberg-Weyl group is clearly insufficient for the production of Gaussian 2-designs.

\section{The symplectic group, and the existence of Gaussian 2-designs} \label{sec:sp}

Each HW-covariant design contains only a subset of the Gaussian states.  Since its elements are related by translation, they all have the same shape in phase space, inherited from the fiducial $\ket{\psi_0}$ (which, without loss of generality, we may take to be a ``squeezed vacuum'' state with $\expect{x} = \expect{p} = 0$).  We need a richer ensemble, containing states with different shapes.  To get it, we observe that every squeezed vacuum state can be obtained from the vacuum state by applying a particular \emph{linear symplectic} transformation.  These are linear transformations on phase space that preserve symplectic area $\diff x\diff p$, and they form the symplectic group Sp$(1,\reals)$.  It is isomorphic to the special linear group SL$(2,\reals)$, whose fundamental representation is the $2\times2$ real matrices with unit determinant \cite{Lang85}.  This 3-parameter Lie group is the natural symmetry group of squeezed vacuum states, on which it acts transitively, and it will be central to our discussion.

Sp$(1,\reals)$, which we will denote by Sp for convenience, contains \emph{all} the linear transformations of phase space (which means all the transformations that preserve Gaussianity).  This includes:
\begin{enumerate}
\item rotations (``elliptic transformations''), of the form $R(\theta) = \begin{pmatrix} \cos\theta & \sin\theta \\ -\sin\theta & \cos\theta \end{pmatrix}$,
\item squeezings (``hyperbolic transformations''), of the form $U(u) = \begin{pmatrix} u^{-1/2} & 0 \\ 0 & u^{1/2} \end{pmatrix}$,
\item shearings (``parabolic transformations''), of the form $V(v) = \begin{pmatrix} 1 & 0 \\ -v & 1 \end{pmatrix}$.
\end{enumerate}
Rotations, squeezings, and shearings each form a 1-parameter subgroup of Sp, but these subgroups are \emph{not} normal.  In fact, the projective symplectic group is simple (has no nontrivial normal subgroups), so it cannot be factored as a group.  However, if viewed as a 3-parameter manifold, it \emph{can} be factored, using the Iwasawa decomposition \cite{IwasawaAM49}, usually written in the notation $\mathrm{Sp} = KAN$, which identifies each element $g$ of a group $G$ with three elements of subgroups $K,A,N$: $g \sim \{k,a,n\}$.
The Iwasawa decomposition writes Sp as a direct product of the three sets $K = \{k\}$, $A = \{a\}$, and $N = \{n\}$.  The group element corresponding to $\{k,a,n\}$ is just the composition (matrix product) of the three subgroup elements, $g = kan$, and the three subgroups $K,A,N$ are precisely the ones listed above -- rotations ($K$), squeezings ($A)$, and shearings ($N$).

Since Sp is transitive on squeezed vacuum states, any squeezed vacuum state can be obtained by applying some $g\in$ Sp to the $\ket{0}$ state.  Adding the generators of the Heisenberg-Weyl group (whose action on phase space is affine, not linear) yields the \emph{affine symplectic group}.  This semidirect product group, HWSp = $\mathrm{HW} \rtimes \mathrm{Sp}(1,\reals)$ is transitive on \emph{all} Gaussian states.  So there is a unique HWSp-covariant design $\cE_\mathrm{Sp}$ containing \emph{all} Gaussian states, weighted by a HWSp-invariant measure, which has the potential to form a Gaussian 2-design.

\subsection{The affine symplectic group is irreducible on the symmetric subspace of $L^2(\reals)^{\otimes 2}$}

It's fairly straightforward to show that HWSp is irreducible on $\cH_{\mathrm{symm}}^{(2)}$ -- which would appear to imply that $\cE_\mathrm{Sp}$ is a 2-design.  An element of HWSp is a projective linear transformation on phase space, and is completely characterized by its action on the position ($\bX$) and momentum ($\bP$) operators, which form a symplectic vector
\begin{equation*}
\vec{\bZ} = \pmat{ \bX \\ \bP }.
\end{equation*}
Heisenberg-Weyl transformations act on $\vec{\bZ}$ additively,
\begin{equation*}
h_{x,p}\left[\pmat{ \bX \\ \bP }\right] = \pmat{ \bX+x \\ \bP+p },
\end{equation*}
while symplectic transformations (represented as $2\times 2$ matrices with unit determinant) act on $\vec{\bZ}$ by matrix multiplication:
\begin{equation*}
s_{a,b,c,d}\left[\pmat{ \bX \\ \bP }\right] = \pmat{a&b\\c&d} \pmat{ \bX \\ \bP } = \pmat{a\bX+b\bP \\ c\bX+d\bP}.
\end{equation*}
So an arbitrary element of HWSp acts on $\vec{\bZ}$ as
\begin{equation*}
\pmat{ \bX \\ \bP }\to \pmat{a\bX+b\bP + x \\ c\bX+d\bP + p}.
\end{equation*}
The 2-copy tensor product action is thus
\begin{equation*}
\pmat{ \bX_1 \\ \bP_1 \\ \bX_2 \\ \bP_2 }\to \pmat{a\bX_1+b\bP_1 + x \\ c\bX_1+d\bP_1 + p \\ a\bX_2+b\bP_2 + x \\ c\bX_2+d\bP_2 + p},
\end{equation*}
so its action on $\bX_\pm$ and $\bP_\pm$ is
\begin{equation}
\pmat{ \bX_+ \\ \bP_+ \\ \bX_- \\ \bP_- }\to \pmat{a\bX_+ +b\bP_+ + \sqrt2x \\ c\bX_+ + d\bP_+ + \sqrt2p \\ a\bX_- + b\bP_- \\ c\bX_- + d\bP_-}. \label{eq:HWSpaction}
\end{equation}
Eq. \ref{eq:HWSpaction} shows that, while the Sp subgroup acts faithfully on \emph{both} factors, the Heisenberg-Weyl subgroup acts faithfully on $L^2(x_+)$ but trivially on $L^2(x_-)$.  Irreducibility follows from three observations.
\begin{enumerate}
\item HWSp contains HW as a subgroup, and HW is irreducible on $L^2(x_+)$, so HWSp is irreducible on this factor.
\item HWSp contains a Sp subgroup that acts faithfully on $L^2(x_-)$.  Sp contains an SO$(2)$ subgroup, represented as rotations on phase space.  These are generated by the harmonic oscillator Hamiltonian $H = \bX_-^2 + \bP_-^2$, so the irrep spaces of the SO$(2)$ representation are its 1-dimensional eigenspaces $\{\ket{n}\}$.  The irrep spaces of Sp on $L^2(x_-)$ must be coarse-grainings of the SO$(2)$ irrep spaces, so they are direct sums of harmonic oscillator eigenspaces.
\item Sp also contains squeezing transformations, which act as 
\begin{equation*}
\pmat{ \bX \\ \bP }\to \pmat{u^{-1/2}\bX \\ u^{1/2}\bP}.
\end{equation*}
Any nontrivial squeezing transformation has nonzero matrix elements between $\ket{0}$ and $\ket{2n}$ for every $n$ -- which is to say that it maps $\ket{2n}\to\ket{0}$ and vice-versa with some amplitude.  Squeezing transformations therefore mix together all the even number states.  They span the positive-parity subspace of $L^2(x_-)$, which therefore has no Sp-invariant subspaces.
\end{enumerate}
Observations 2 and 3 imply that HWSp is irreducible on the positive-parity subspace of $L^2(x_-)$.  Together with Observation 1, this implies that HWSp is irreducible on the symmetric subspace of $L^2(\reals)^{\otimes 2}$.

This proof is entirely nonconstructive; it does not even suggest what is the HWSp-invariant measure over Gaussian states.  Since HWSp is noncompact, the measure will not be normalizable.  This doesn't necessarily pose problems -- the same is true of the Heisenberg-Weyl group, but the coherent state POVM is perfectly well-defined and physically meaningful.  However, when we attempt to \emph{construct} the 2-design whose existence seems to be implied by the irreducibility of HWSp, we shall see that things get rather confused.

\section{Explicit constructions, and the nonexistence of Gaussian 2-designs} \label{sec:no}

Let us try to construct a Gaussian 2-design $\cE$.  To keep things clear and simple, we will not try (yet) to compute the invariant measure over HWSp.  Instead, we simply assume that $\cE$ is invariant under the Heisenberg-Weyl subgroup, and under the SO$(2)$ subgroup of phase space rotations.  These invariances imply:
\begin{enumerate}
\item The ensemble-average state will be maximally mixed on the $L^2(x_+)$ factor (which can therefore be ignored) because HW is irreducible on that factor.
\item $\cE$ comprises Heisenberg-Weyl \emph{orbits}, each defined by (i) applying a specific linear symplectic transformation $g\in$ Sp to $\ket{0}^{\otimes 2}$ to get a Gaussian $\ket{\psi_g}^{\otimes 2}$, and (ii) applying all $T_{x,p}\in\mathrm{HW}$ to $\ket{\psi_g}^{\otimes2}$.  Each orbit is a tensor product, between a Heisenberg-Weyl covariant 1-design on $L^2(x_+)$, and a single squeezed vacuum state $\proj{\psi_g}$ on $L^2(x_-)$.  This is true because HW acts trivially on the $L^2(x_-)$ factor, while Sp acts faithfully on the $L^2(x_-)$ factor.
\end{enumerate}
Now, each squeezed vacuum state $\ket{\psi_g}$ on $L^2(x_-)$ can be described by its degree of squeezing $s$ (see Eq. \ref{eq:sdef}), and the angle $\theta$ in phase space between its major axis and the $x$-axis.  The HWSp-invariant measure over states will be \emph{something} of the form
\begin{equation}
\mu(s,\theta)\diff s\diff\theta.
\end{equation}
So, when we calculate
\begin{equation*}
\Avg_{\ket{\psi}\in\cE}\left(\proj{\psi}^{\otimes 2}\right),
\end{equation*}
we will be adding up a lot of 2-mode Gaussian states.  And, thanks to the first two points above, we know that as long as the ensemble is Heisenberg-Weyl-covariant, 
\begin{equation*}
\Avg_{\ket{\psi}\in\cE}\left(\proj{\psi}^{\otimes 2}\right) \propto \Id_{+}\otimes\Avg_{\ket{\psi_g}\in\cE}\proj{\psi_g}_{-},
\end{equation*}
where for each $\ket\psi$ in $\cE$, the corresponding $\ket{\psi_g}$ is a squeezed vacuum state with the same shape (but $\expect{x}=\expect{p}=0$).  $\cE$ is a 2-design if and only if
\begin{equation*}
\Avg_{\ket{\psi_g}\in\cE}\proj{\psi_g} \propto \Pi = \frac{(\Id+P)_{-}}{2}.
\end{equation*}
The R.H.S. is the projector on the symmetric subspace.  In the ``number basis'' $\{\ket{k}:k=0,1,2,\ldots\}$ (eigenstates of $H = \bX^2 + \bP^2$), parity acts as $P\ket{k} = (-1)^k\ket{k}$, so the projector onto positive parity states is
\begin{equation*}
\Pi = \sum_{k=0}^\infty{\proj{2k}}.
\end{equation*}
But this operator cannot be built as a sum of Gaussian projectors $\proj{\psi_g}$, as we will now show by computing the diagonal matrix entries of a single $\proj{\psi_g}$ -- where $g\sim(s,\theta)$ -- in the number basis.

\subsection{Detailed calculation of overlaps}

Suppose that $\ket{\psi_g} = \ket{\psi_{\theta,s}}$ is an arbitrary squeezed vacuum state, and let
\begin{equation*}
p_k(\theta,s) = |\braket{k}{\psi_{\theta,s}}|^2.
\end{equation*}
Now, $\ket{\psi_{\theta,s}}$ is parity-symmetric, and if $k$ is odd then $\ket{k}$ is parity-antisymmetric. So $p_k=0$ for odd $k$.  Furthermore, $\proj{k}$ is invariant under phase space rotations, so $\theta$ has no impact on $p_k$.  We thus write it as $p_k(s)$, and assume without loss of generality that $\ket{\psi_s}$ is squeezed along the $x$ axis.  The corresponding wavefunction is 
\begin{equation*}
\psi_s(x) = \braket{x}{\psi_{\theta=0,s}} = (\pi s^2)^{-1/4} e^{-x^2/2s^2},
\end{equation*}
yielding squeezed quadratures
\begin{equation}
\pmat{ \Delta x^2 & \Delta xp \\ \Delta xp & \Delta p^2 } = \frac12\pmat{ s^2 & 0 \\ 0 & 1/s^2}. \label{eq:sdef}
\end{equation}

The wavefunction for a number state $\ket{k}$ is the product of a Hermite polynomial and the ground-state Gaussian wavefunction:
\begin{equation*}
\phi_k(x) = \braket{x}{k} = \frac{H(k,x) e^{-x^2/2}}{(\pi)^{1/4}\sqrt{2^k k!}},
\end{equation*}
where 
\begin{equation*}
H(k,x) = \left[ \left(\frac{\partial}{\partial \, t}\right)^k e^{2xt-t^2} \right]_{t=0}.
\end{equation*}
Using these wavefunctions, for $k$ even,
\begin{eqnarray}
\sqrt{p_k(s)} &=& |\braket{k}{\psi_s}| \nonumber \\ 
&=& \int{\phi_k(x)\psi_s(x)\diff x} \nonumber \\
&=& \frac{1}{\sqrt{\pi 2^{k} k! s}} \left(\frac{\partial}{\partial \, t}\right)^k \left[ \int{ e^{2xt-t^2-x^2(1+1/s^2)/4} \diff x} \right]_{t=0} \nonumber \\
&=& \frac{1}{\sqrt{\pi 2^{k} k! s}} \left( \frac{\partial}{\partial \, t}\right)^k \left[ \frac{s\sqrt{2\pi}}{\sqrt{s^2+1}}e^{t^2(s^2-1)/(s^2+1)} \right]_{t=0} \nonumber \\
&=& \sqrt{\frac{s}{2^{k-1} k! (s^2+1)}} \left[ \left(\frac{s^2-1}{s^2+1}\right)^{k/2} \frac{k!}{(k/2)!} \right]\nonumber \\
&=& \sqrt{\frac{2^{k+1} s (s^2-1)^k}{\pi k! (s^2+1)^{k+1}}}\Gamma\left(\frac{k+1}{2}\right) \nonumber
\end{eqnarray}
Squaring this expression and rewriting $k! = \Gamma(k+1)$ yields
\begin{equation*}
p_k = \frac{2^{k+1}\Gamma\left(\frac{k+1}{2}\right)^2s}{\pi\Gamma(k+1)(s^2+1)}\left(\frac{s^2-1}{s^2+1}\right)^k.
\end{equation*}
We now use the identity
\begin{equation*}
\Gamma(2z) = \frac{1}{\sqrt{2\pi}} 2^{2z-1/2}\Gamma(z)\Gamma(z+1/2)
\end{equation*}
to rewrite the denominator using
\begin{equation*}
\Gamma(k+1) = \pi^{-1/2}2^k\Gamma\left(\frac{k}{2}+\frac12\right)\Gamma\left(\frac{k}{2}+1\right),
\end{equation*}
which yields
\begin{equation*}
p_k = \frac{2\Gamma\left(\frac{k+1}{2}\right)s}{\sqrt{\pi}\Gamma\left(\frac{k}{2}+1\right)(s^2+1)}\left(\frac{s^2-1}{s^2+1}\right)^k.
\end{equation*}
Finally, we observe that $p_k$ is symmetric under $s \to 1/s$ -- as it should be -- and with this in mind we rewrite it in terms of the squeezed vacuum state's energy
\begin{equation*}
E = \frac{\expect{X^2}+\expect{P^2}}{2} = \frac{\Delta x^2 + \Delta p^2}{2} = \frac14\left(s^2 + \frac{1}{s^2}\right).
\end{equation*}
Solving for $s$ gives $s = \sqrt{2E\pm\sqrt{4E^2-1}}$, and therefore
\begin{eqnarray*}
\frac{s}{s^2+1} &=& \frac{1}{\sqrt{4E+2}} \\
\frac{s^2-1}{s^2+1} &=& \sqrt{\frac{2E-1}{2E+1}}.
\end{eqnarray*}
This yields our final result:
\begin{eqnarray} 
p_k(E) &\equiv& |\braket{k}{\psi_{s,\theta}}|^2 \nonumber \\
&=& \sqrt{\frac{2}{\pi}}\frac{\Gamma\left(\frac{k}{2}+\frac12\right)}{\Gamma\left(\frac{k}{2}+1\right)}\cdot\frac{1}{\sqrt{2E+1}}\cdot\sqrt{\frac{2E-1}{2E+1}}^k \label{eq:pk}
\end{eqnarray}
(for even $k$; $p_k$ vanishes when $k$ is odd.)

\subsection{There are no Gaussian 2-designs} \label{sec:fail}

Let us now consider the implications of Eq.~(\ref{eq:pk}).  Our proposed ensemble will contain Gaussian states with some measure $f(E)\diff E$ over $E$, which has not yet been defined.  But states with each value of $E\in[1,\infty)$ will contribute to the average of $\proj{\psi}$ on $L^2(x_-)$, according to
\begin{eqnarray}
\Avg_{\psi\in\cE}\left(\proj{\psi}\right) &=& \int{\left(\sum_{k=0}^{\infty}{p_k(E)\proj{2k}}\right)f(E)\diff E} \nonumber \\
&=& \sum_{k=0}^{\infty}{\int{p_k(E)f(E)\diff E}\proj{2k}}. \label{eq:Eintegral}
\end{eqnarray}
This equation is entirely correct as long as the measure over Gaussian states is invariant under rotations in phase space (in which case the off-diagonal elements vanish by symmetry); otherwise it is only correct for the diagonal elements, which is still sufficient to demonstrate a problem.

Now, $p_k$ splits into a factor which depends on the squeezing (energy) of the Gaussian state,
\begin{equation*}
\frac{1}{\sqrt{E+1}}\cdot\sqrt{\frac{E-1}{E+1}}^k,
\end{equation*}
and a factor that is completely independent of the state,
\begin{equation*}
\frac{\Gamma\left(\frac{k}{2}+\frac12\right)}{\sqrt{2\pi}\Gamma\left(\frac{k}{2}+1\right)},
\end{equation*}
so
\begin{equation*}
\Avg_{\psi\in\cE}\left(\proj{\psi}\right) = \sqrt{\frac{2}{\pi}}\sum_{k=0}^{\infty}{\frac{\Gamma\left(k+\frac12\right)}{\Gamma(k+1)}w_k(E)\proj{2k}},
\end{equation*}
where $w_k(E)$ contains all the $E$-dependence of this average state, and is given by
\begin{equation*}
w_k(E) = \int{\frac{1}{\sqrt{E+1}}\left(\frac{E-1}{E+1}\right)^k f(E)\diff E}.
\end{equation*}
Now, without saying anything about the measure over $E$, we can see a problem.  The integrand of $w_k(E)$ is a strictly decreasing function of $k$ -- for \emph{every} finite value of $E$ -- so $w_k(E)$ is strictly nonincreasing with $k$.  But the $E$-independent coefficient is \emph{also} strictly decreasing; the first few values of
\begin{equation*}
\frac{\Gamma\left(k+\frac12\right)}{\Gamma(k+1)}
\end{equation*}
are
\begin{equation*}
\sqrt{\pi},\ \frac{\sqrt{\pi}}{2},\ \frac{3\sqrt{\pi}}{8},\ \frac{5\sqrt{\pi}}{16},\ \frac{35\sqrt{\pi}}{128},\ \frac{63\sqrt{\pi}}{256},\ \frac{231\sqrt{\pi}}{1024}, \ldots
\end{equation*}
and by $k=2$ it is already within $1\%$ of its asymptotic value
\begin{equation*}
\lim_{k\to\infty}{\frac{\Gamma\left(k+\frac12\right)}{\Gamma(k+1)}} = \frac{1}{\sqrt{k+1/4}}.
\end{equation*}
So the average state is bounded above by the product of an operator with strictly decreasing eigenvalues and
\begin{equation}
\sum_{k=0}^\infty{\frac{\proj{2k}}{\sqrt{k+1/4}}}.\label{eq:sqrtop}
\end{equation}
This is \emph{not} equal to the projector onto the positive-parity subspace -- which would have a flat spectrum.  So it is \emph{not} possible to build a 2-design out of Gaussian states.  

In fact, it's not even possible to get close.  The best we can do is to choose an ensemble of very highly squeezed states, so that $E\gg 1$ for all states.  The $E$-dependent terms in Eq.~(\ref{eq:pk}) go to 1, and we are left with something proportional to the operator in Eq.~(\ref{eq:sqrtop}) -- which is not in any sense ``close'' to the projector that would signify a 2-design.

\section{So what is going on?}\label{sec:discuss}

The previous sections appear to present a contradiction.  On one hand, HWSp is irreducible on the symmetric subspace of $L^2(\reals)\otimes L^2(\reals)$, so Schur's Lemma seems to imply that a HWSp-invariant ensemble of Gaussian states must be a 2-design.  On the other hand, the explicit calculation of the last section makes it clear that no mixture of Gaussian states can yield a 2-design.

The resolution to this tension is to observe that Schur's Lemma states only that \emph{if} an operator $X$ on an irreducible representation space of $G$ commutes with every element of the representation $\{T_g\}$, then $X \propto \Id$ on that space.  It does \emph{not} guarantee the existence of $X$.  If the $G$-twirled operator
\begin{equation*}
\rho = \int{T_g\proj{\psi}T^\dagger_g\diff g}
\end{equation*}
exists, then (by construction) it must commute with every $T_g$ (and then Schur's Lemma completes the proof).  But it may not exist!  When $G$ is not compact, the integral may or may not converge.  When it does not converge -- as turns out to be the case here -- the $G$-twirled state does not exist, and Schur's Lemma is not applicable.

The symplectic group provides a particularly interesting (and -- to a physicist -- irritating) case, because:
\begin{enumerate}
\item Sp \emph{does} have a well-defined Haar measure (see next section),
\item there is in fact an Sp-invariant operator on $L^2(\reals)$ (the identity operator) and a HWSp-invariant operator on $L^2(\reals)\otimes L^2(\reals)$ (the projector on the symmetric subspace),
\item $G$-twirling frequently \emph{does} converge, even for noncompact groups $G$ (e.g., when $G = \mathrm{HW}$, as we showed in Section \ref{sec:h}).
\end{enumerate}
The last point is illustrated by the Heisenberg-Weyl group.  While it is noncompact, that does not prevent us from HW-twirling the vacuum state to get
\begin{equation*}
\int_{\mathrm{HW}}{T_{x,p}\proj{0}T_{x,p}^\dagger\diff x\diff p} \propto \Id.
\end{equation*}
Actually, a rather subtle trick is required to make this integral converge.  If we define a sequence of partial integrals over $|x+ip|<r$ for $r = 1,2,4,8,16,\ldots$, then the sequence of partial integrals is not Cauchy -- i.e., it does not converge to an operator in $\mathcal{B}(L^2(\reals))$.  However, if we project these partial integrals onto \emph{any} finite subspace of $L^2(\reals)$, then the sequence of projections converges to $\Id$ on that subspace.

In contrast, the Sp-twirling integral diverges on \emph{every} finite subspace.  Let us show this explicitly by (finally) deriving the Sp-invariant measure over Gaussian states.

\section{The symplectically-invariant measure}

Although Sp is a 3-parameter Lie group, we are really only concerned with its action on the 2-parameter manifold of Gaussian squeezed vacuum states.  So we do not need the whole group, nor do we need to calculate the Haar measure over the entire group.  However, we need a subgroup that is transitive on squeezed vacuum states.  The 2-parameter parabolic subgroup of Sp generated by squeezings $U$ and shearings $V$ (see Sec.\ref{sec:sp}) is exactly such a group \cite{SimonPRA88}.

An arbitrary element of the parabolic subgroup is given by $S(u,v) = V(v)U(u)$.  Composition of two elements is given by
\begin{equation*}
S(u',v')S(u,v) = S(u'u,u'v+v').
\end{equation*}
Viewed as a left action, this implies that the measure $\mu(u,v)$d$u$d$v$ transforms under $S(u',v')$ as
\begin{equation*}
\mu(u,v) \mapsto \mu(u'u,u'v+v') |u'^2|,
\end{equation*}
where $|u'^2|$ is the Jacobian of the transformation.  Let us choose normalization so that the measure equals $\diff u\diff v$ at the identity element ($u=1$, $v=0$), so $\mu(1,0)=1$.  Then we have $\mu(u',v')u'^2=1$, so the left invariant Haar measure is 
\begin{equation}\label{eq:uvmeas}
\mu = \frac{1}{u^2} \mathrm{d}u \mathrm{d}v.
\end{equation}
Note that, although Sp$(1,\mathbb{R})$ is unimodular (its left and right invariant measures are equal), the parabolic subgroup is not; it has modulus $1/|u'|$ when $S(u',v')$ acts on the right.

Since this group is transitive on squeezed vacuum states, we can parameterize them using $(u,v)$ as well.  To do so, we observe that a squeezed vacuum state is completely specified by a real, symmetric, positive definite matrix $G$ with unit determinant (the state's covariance matrix is $\frac12G^{-1}$).  $G$ defines the state's Wigner function on phase space (parameterized by $Z=[x,p]^\mathrm{T}$) as
\begin{equation*}
W_G(Z) = \frac{1}{\pi} \exp(-Z^\mathrm{T} G Z),
\end{equation*}
We can apply a symplectic transformation contragradiently to $Z$, as $Z\mapsto S^{-1}Z$, or we can apply it to $G$ as $G\mapsto (S^{-1})^\mathrm{T} G S^{-1}.$  So, to find the matrix $G(u,v)$ for a specific squeezed state, we apply squeezing and shearing transformations to the vacuum state ($G=\Id$), obtaining
\begin{eqnarray*}
G(u,v) &=& V(-v)^\mathrm{T}U(1/u)^\mathrm{T}U(1/u)V(-v) \nonumber \\
&=& \begin{pmatrix} u+v^2/u & v/u \\ v/u & 1/u \end{pmatrix}
\end{eqnarray*}
for a general squeezed vacuum state.  We could reach the same state by squeezing and then rotating,
\begin{equation*}
G(s,\theta) = R(-\theta)^\mathrm{T} U(s^2)^\mathrm{T} U(s^2) R(-\theta),
\end{equation*}
provided that 
\begin{eqnarray*}
u
&=& \frac{s^2}{s^4\cos^2(\theta)+\sin^2(\theta)} ,\\
v
&=& \frac{(s^4-1)\sin(\theta)\cos(\theta)}{s^4\cos^2(\theta)+\sin^2(\theta)}.
\end{eqnarray*}
The measure in Eq.~(\ref{eq:uvmeas}) becomes
\begin{equation*}
\mu = \frac{2|1-s^4|}{s^3} \mathrm{d}s \mathrm{d}\theta,
\end{equation*}
and if we substitute $s = \sqrt{2E\pm\sqrt{4E^2-1}}$, we find (after some algebra)
\begin{equation}
\mu = 4\diff E\diff\theta. \label{eq:HWSpmeasure}
\end{equation}
The average energy of the invariant ensemble is divergent, and -- since every energy contributes equally -- it is dominated by states with arbitrarily large squeezing.  Furthermore, every nonzero matrix element in the ensemble-average state defined by Eq. \ref{eq:Eintegral} diverges.  This explains (and resolves) the apparent contradiction with Schur's Lemma:  there \emph{is} no ensemble-average positive operator, since the integral diverges irreconcilably.  Since the average doesn't exist, it need not be proportional to an irrep projector.  

Note that the measure defined in Eq. \ref{eq:HWSpmeasure} is perfectly well-defined and well-behaved.  Some functions have well-defined averages over it, but some (notably the positive-operator-valued function $\proj{\psi}^{\otimes 2}$) do not.  Consider (as a simple example of the same phenomenon), the standard Lebesgue measure $\diff x$.  Some functions, such as $f(x) = (x^2+1)^{-1}$, have well-defined integrals over $\diff x$.  Others, such as $f(x) = x^{-1}$, do not.  Although physicists can often use tricks to evaluate ill-defined integrals (e.g., setting $\int_{-\infty}^{\infty}{x^{-1}\diff x}=0$ by symmetry), some integrals over $\diff x$ are truly ill-defined -- like that of $f(x)=1$.  For squeezed Gaussians, the ensemble-average state is such a case.

\section{Conclusion}

The main result of this paper is, of course, that Gaussian 2-designs do not exist.  There \emph{is} a natural measure over all Gaussian states, which is left-invariant under the entire affine symplectic group, but neither this measure nor any other can yield a 2-design.  This is not, of course, to say that there are no 2-designs for $L^2(\reals)$ -- just that they cannot be built of Gaussian states.

Equation \ref{eq:sqrtop} shows that Gaussian ensembles can't even get arbitrarily close to the 2-design condition.  This is significant for state and process tomography, since the spectral non-uniformity of Eq. \ref{eq:sqrtop} means that any Gaussian POVM will be less than optimally sensitive to variations in certain matrix elements.  \emph{However}, this by no means implies that squeezing does not help in quantum tomography.  While squeezed states do not form a 2-design, our analysis has shown that they come far closer to doing so than do coherent states.  Squeezed measurements promise major improvements in the statistical efficiency with which non-classical states can be reconstructed -- just not quite as much improvement as (in principle) could be gotten with non-Gaussian states.

In particular, we note that the eigenvectors of position, momentum, and $\mathbf{X}\cos\theta + \mathbf{P}\sin\theta$ (for uniformly distributed $\theta$), which are measured in the theoretical idealization of heterodyne quantum state tomography \cite{VogelPRA89}, can be seen as infinitely squeezed Gaussian states (or, more precisely, as a $s\to\infty$ limit of Gaussian states).  This ``heterodyne POVM'', rotationally invariant and dominated by infinitely squeezed states, is essentially identical to the HWSp-invariant measure.  So while it is still rather far from being a 2-design, it is as close as \emph{any} Gaussian POVM can be -- and far more uniformly sensitive than the coherent-state POVM.

We still find the nonexistence of Gaussian 2-designs to be quite ``curious''; the representation-theoretic argument given in Section \ref{sec:sp} had us convinced for quite some time that they \emph{did} exist.  The counterargument is unique to noncompact groups and infinite-dimensional Hilbert spaces.  It also demonstrates a fundamental difference between the symplectic phase space structure on finite Hilbert spaces, where irreducible representations \emph{do} reliably yield designs, and $L^2(\reals)$ where, as we have shown, irreducibility does not necessarily imply that $\proj{\psi}^{\otimes 2}$ can be integrated over the group.  

\begin{acknowledgments}
P.S.T. acknowledges useful discussion with S. Bartlett, A. Harrow, and J. Repka, as well as support from JSPS Research Fellowships for Young Scientists, JSPS KAKENHI (20549002) for Scientific Research (C).  R.B.K. is supported by LANL's LDRD program.
\end{acknowledgments}

\bibliographystyle{apsrev}

\bibliography{G2design}

\end{document}